\documentclass[%
superscriptaddress,
 amsmath,amssymb,
 aps,
prb,twocolumn,longbibliography
]{revtex4-1}

\usepackage{graphicx}% Include figure files
\usepackage{mathtools}
\usepackage{amsmath}
\usepackage[caption = false]{subfig}
\usepackage{amsthm}
\usepackage{dcolumn}% Align table columns on decimal point
\usepackage{bm}% bold math
\usepackage[usenames,dvipsnames]{xcolor}
\usepackage[colorinlistoftodos]{todonotes}
\usepackage[colorlinks=true,linkcolor=blue,urlcolor=blue,citecolor=blue]{hyperref}
\usepackage{bbold} % for Identity symbol

\newcommand{\Id}{\mathbb{1}}
\newcommand{\bra}[1]{\langle #1 |}
\newcommand{\ket}[1]{| #1 \rangle}

\begin{document}

\title{Almost-conserved operators in nearly many-body localized systems}
\author{Nicola Pancotti}%
\affiliation{%
Max-Planck-Institut f\"ur Quantenoptik, Hans-Kopfermann-Str. 1, 85748 Garching, Germany
}%

\author{Michael Knap}%
\affiliation{%
Department of Physics and Institute for Advanced Study, Technical University of Munich,
85748 Garching, Germany
}%
\author{David A. Huse}%
\affiliation{%
Physics Department, Princeton University, Princeton, New Jersey 08544, USA
}%
\author{J. Ignacio Cirac}%
\affiliation{%
Max-Planck-Institut f\"ur Quantenoptik, Hans-Kopfermann-Str. 1, 85748 Garching, Germany
}%
\author{Mari Carmen Ba\~nuls}%
\affiliation{%
Max-Planck-Institut f\"ur Quantenoptik, Hans-Kopfermann-Str. 1, 85748 Garching, Germany
}%
\date{\today}

\begin{abstract}
We construct almost conserved local operators, that possess a minimal commutator with the Hamiltonian of the system, near the many-body localization transition of a one-dimensional disordered spin chain. We collect statistics of these slow operators for different support sizes and disorder strengths, both using exact diagonalization and tensor networks. Our results show that the scaling of the average of the smallest commutators with the support size is sensitive to  Griffiths effects in the thermal phase and the onset of many-body localization. Furthermore, we demonstrate that the probability distributions of the commutators can be analyzed using extreme value theory and that their tails reveal the difference between diffusive and sub-diffusive dynamics in the thermal phase. 

\end{abstract}

\maketitle

\section{\label{sec:level1}Introduction}

Isolated interacting quantum systems can undergo an unconventional transition from an ergodic thermal phase in which the system approaches thermal equilibrium irrespective of the initial conditions to a many-body localized (MBL) phase in which it does not.\cite{BAA, Gornyi, nhreview, altman_universal_2015} There exists firm evidence, from numerical,\cite{oh, znidaric2008} rigorous mathematical,\cite{jzi} and experimental work,\cite{schreiber2015observation,demarco2015,smith2015,Bordia16,Hild16,bordia_periodically_2017,lueschencrit16,bordia_probing_2017} that the MBL phase is realized in strongly disordered and interacting quantum systems. Recently, the MBL transition between the thermal and the MBL phase has been started to be investigated using general arguments,\cite{grover2014, Agarwal, Gopalakrishnan15, chandran_finite_2015, gopalakrishnan_griffiths_2016, agarwal_rr_16, khemani_critical_2017} mean-field theory,\cite{gn} and renormalization-group schemes~\cite{VHA, PVPtransition, Zhang_2016}. Since the MBL transition is inherently dynamical and can occur at arbitrary energy density, it is beyond the scope of a conventional thermodynamic description. 

A possible strategy to explore MBL is to directly study properties of the Hamiltonian.
% instead of observables. 
In particular, a phenomenological description of the Hamiltonian in the MBL phase using an extensive set of local integrals of motion (LIOMs) has been proposed.\cite{huse_phenomenology_2014, serbyn_local_2013} This approach explains, for instance, the dephasing~\cite{serbyn_interferometric_2014, bahri_localization_2015, Chandran_2016} and entanglement dynamics~\cite{znidaric2008, bardarson_unbounded_2012} in the localized phase. Moreover, LIOMs have been constructed explicitly for some models using analytical techniques,~\cite{ros15liom} exact  diagonalization~\cite{rademaker16liom, chandran_constructing_2014,OBrien_2016, He_2016, thomson17flow, Goihl_2017}, stochastic methods,~\cite{inglis_accessing_2016} and tensor networks.~\cite{chandran15specTN, pekker17prb, pollmann_2016, wahl17prx} Upon reducing the disorder, LIOMs become more and more extended in space and eventually cease to exist at the MBL transition.

Here, we take another route and study the dynamics of a disordered and interacting Heisenberg spin chain by constructing almost conserved local operators with finite support, which minimize the commutator with the system Hamiltonian.\cite{Kim_2015}  Deep in the MBL phase the slow operators resemble LIOMs. However, our procedure can be directly extended to the thermal phase as well. The slow operators can potentially be used to study dynamical properties, because the value of the commutator gives a lower bound on the thermalization times of the corresponding operator.\cite{Kim_2015} We compute the slowest operators for different support sizes and disorder strengths, using both exact diagonalization and tensor networks.\cite{verstraete08num,Schollwoeck201196} 
Our work, rather than focusing on the phase transition, it aims at characterize the sub-diffusive region in its vicinity. Specifically, we collect statistics from numerous disorder realizations and show that not only the mean value of the smallest commutator, but also the probability distributions are sensitive to the underlying phase. The \emph{tails} of the distributions are affected by the appearance of rare Griffiths regions,\cite{Agarwal, agarwal_rr_16}  which act as bottlenecks for transport and are argued to provide a simple description for the slow sub-diffusive dynamics in the thermal phase near the MBL transition.~\cite{BarLev_Absence_2015, Agarwal, znidaric_diffusive_2016-1} We use extreme value theory (EVT) to analyze the tails of the distribution and find that it is well described by the generalized extreme value distribution. \cite{Haan_EVT} Moreover, we discuss how the presence of rare regions may affect the asymptotic behavior of such distributions. Our results demonstrate that finding slow operators provides access dynamical quantities 
without resorting to the time evolution of a particular state.

 The paper is organized as follows.
 In section \ref{sec:model} we specify our model. 
  Section \ref{sec:methods} describes the slow operator method, and presents the fundamental ideas of EVT employed in our analysis.
  The results for the structure of the slow operators are presented in section \ref{sec:structure}, the average values of their commutators in the different phases are shown in section \ref{sec:avg}, and the statistical analysis of the probability distributions using EVT is presented in section
   \ref{sec:evt}.
   Finally, in section \ref{sec:discussion} we summarize our findings and discuss potential extensions of our work.

\section{Model}
\label{sec:model}

We consider the non-integrable spin-1/2 Heisenberg Hamiltonian with random magnetic field,
\begin{equation} \label{hamiltonian}
H = \sum_{i} J \left( S^x_i S^x_{i+1} + S^y_i S^y_{i+1} + S^z_i S^z_{i+1} \right) + h_i S^z_i  , 
\end{equation}   
where $S_i^{\alpha}$ are spin-1/2 operators, and the values of the transverse field $h_i$ are randomly chosen from the uniform distribution in the interval $\left[ -h, h \right]$, where $h$ is the disorder strength.
In the following, we set $J=1$. This model exhibits a phase transition from a thermal to a MBL phase at $h_c \sim 4$.\citep{Luitz_2015, oh}

In the MBL phase the system can be described by LIOMs, which lead to an emergent integrability. The corresponding effective Hamiltonian can be expressed as~\cite{huse_phenomenology_2014, serbyn_local_2013}
\begin{equation}
\small
H = \sum_i \tau_i^z + \sum_{ij} J_{ij} \tau_i^z \tau_j^z + \sum_{n} \sum_{i,j,\{k\}} K^{(n)}_{i\{k\}j} \tau_i^z \tau_{k_1}^z...\tau_{k_n}^z \tau_j^z.
\label{eq:lbits}
\end{equation}
The LIOMs, $\tau_i^z$, commute with each other and with the Hamiltonian. They
are exponentially localized when written in terms of the physical spins $\{ \vec{S}_i \}$. Moreover, the coefficients $J_{ij}$ and $K^{(n)}_{i\{k\}j}$ decay exponentially with their spatial range.

\section{Methods}
\label{sec:methods}

\subsection{Slow operator technique} \label{subsec:slow_op}

Almost conserved quasi-local operators can identify long thermalization time scales. A way to compute such time scales is to search for slow operators that minimize the commutator with the Hamilonian\cite{Kim_2015} (see also Refs. \onlinecite{mierzejewski15, Mierzejewski_2017,lin17slow}). In particular, we may restrict the search to operators $O_M$ with finite support on $M$ consecutive sites. Defining $\mathcal{L}(O_M)$ to be the squared norm of the commutator between the Hamiltonian and the (normalized) operator, the problem reduces to solving the variational minimization
\begin{equation} 
\lambda_M := \min_{O_M;\, \mathrm{tr}{O_M}=0}  \mathcal{L}(O_M)  = \min_{O_M;\, \mathrm{tr}{O_M}=0}  \frac{\|[H,O_M]\|_F^2}{\|O_M\|_F^2},
\label{lambda}
\end{equation}
where $\|A\|_F^2=\mathrm{tr}(A^{\dagger}A)$ is the Frobenius norm, 
and $O_M$ is restricted to be traceless.\footnote{The traceless condition ensures the solution will have no overlap with the trivially commuting identity.} The operator that minimizes \eqref{lambda} is the one evolving the \emph{slowest} in Heisenberg picture at time $t=0$. 
Moreover, for a slightly perturbed infinite temperature state of the form $\rho(O_M) = {\Id}/{Z} + \epsilon O_M$ the thermalization time is lower bounded as $t_{\mathrm{th}}\geq {1}/{\sqrt{\lambda_M}}$.\cite{Kim_2015}

If the system has a conserved extensive quantity, such as the total energy, it naturally gives rise to slow operators. For a translationally invariant system, the Hamiltonian commutes with the sum of all translations of the corresponding local operator, but not with translations restricted to a finite window $M$. Nevertheless, the norm of the latter commutator seems to decrease polynomially with $M$.\cite{Kim_2015}

In our particular case of Hamiltonian \eqref{hamiltonian}, the total polarization in the $z$ direction, $\sum_i \sigma_i^{z}$, 
is conserved for any value of the disorder strength.
The restriction of the sum to $M$ sites, $Z_M=\sum_{i=0}^M \sin \left( \frac{\pi k}{M} \right) \sigma_i^z$,
is thus a natural slow operator for any given support $M$. 
It is easy to check that as the support increases, 
the corresponding commutator decays as $\mathcal{L}(Z_M)\sim{1}/{M^2}$.\cite{Kim_2015}
This sets a reference for comparison with the commutators found by numerically solving the variational problem, Eq. \eqref{lambda}.
\footnote{In general we expect that a numerical minimization for any given support, will find operators slower than the polarization fluctuations, even in the clean system. \cite{Kim_2015}}

For small support size $M$ the optimization can be solved via exact diagonalization. This has been used to show that for model  \eqref{hamiltonian} there exists an extensive 
number of exponentially localized LIOM in the MBL phase 
which are not present in the thermal phase~\cite{OBrien_2016}.
To reach larger supports the search can be restricted 
to operators  $O_M$ of the form of a matrix product operator (MPO). 
Then standard tensor network techniques can be applied to solve this
optimization, and the bond dimension can then be systematically increased
until convergence is achieved.

In this work, we use exact and approximate MPO solutions in order to
collect information on the minimal commutators and the corresponding operators. For different values of the support $M$ and the disorder strength $h$, we compute $\lambda_M$ for several configurations of the random magnetic field (ranging from $2 \cdot 10^3$ for $M=12$ to $10^5$ for $M = 4$). 
In order to simplify the statistical analysis, we choose independent configurations over $M+2$ sites. This is equivalent to choosing non-overlapping windows over an infinite chain. 

To solve the problem numerically, the minimization Eq. \eqref{lambda} can be interpreted as an eigenvalue problem. For fixed support size $M$ and disorder strength $h$ and for a particular disorder realization, there is an effective operator $\mathcal{H}_{\mathrm{eff}}^{(M)}$ acting on the $d^{2M}$-dimensional space with support on the chosen window, such that the (vectorized) operator that minimizes Eq. \eqref{lambda} corresponds to its eigenvector with lowest eigenvalue. Since the optimization is restricted to operators supported on a certain window of size $M$, only terms in the Hamiltonian that overlap with that region will contribute to the commutator. For our nearest-neighbor Hamiltonian this means that $M+2$ sites of the Hamiltonian contribute, which we denote by $H_{M+2}$. The matrix representation of the corresponding commutator, acting on vectorized operators, can be written as
$\mathcal{C}_{M+2}= H_{M+2} \otimes \Id - \Id \otimes H_{M+2}^T$,
and the effective operator can be constructed as
\begin{align}
\bra{O_M} \mathcal{H}_{\mathrm{eff}}^{(M)} \ket{O_M} &:=
\mathrm{tr} \left( [H_{M+2},\tilde{O}_M]^{\dagger} [H_{M+2},\tilde{O}_M]\right )
\nonumber\\
&=\bra{\tilde{O}_M} \mathcal{C}_{M+2} \mathcal{C}_{M+2}^{\dagger} \ket{\tilde{O}_M },
\end{align} 
where $\tilde{O}_M=\Id \otimes O_M \otimes \Id$ and the trace is taken over the space of all $M+2$ sites.

\subsection{Extreme value theory}\label{subsec:evt}

The statistical analysis of the smallest commutator contains relevant information about the phases of the model and the critical region, both in the average scaling of $\lambda_M$ with the support size $M$ and in the probability distributions. In particular, the probability density function (PDF) $p(\lambda_M)$ and the corresponding cumulative density function (CDF), $F(\lambda_M) = \int_{-\infty}^{\lambda_M} p(x) dx $, will be sensitive to the presence of rare regions of atypically large disorder which emerge in the proximity of the critical point.

As we show in this work, these PDFs can be described and analyzed within the mathematical framework of Extreme Value Theory (EVT), a branch of statistics concerned with the 
description of rare events,\cite{Haan_EVT} which is applied to floods, earthquakes or risk in finance and insurance, as well as to several branches of physics, including statistical mechanics and critical phenomena. \citep{Bouchaud_1997, Biroli_2007, Juhasz_2006}

In particular, EVT deals with the asymptotic behavior of the
extreme values of a sample, i.e., with the tails of the distributions. 
Let us consider a random variable $x$ governed by a CDF $F(x)$.
Then EVT ensures that the maxima (equivalently the minima) of samples of $F(x)$, properly normalized and centered, will be governed by a CDF\cite{Haan_EVT}

\begin{equation}\label{eq:gev}
G_{\zeta} \left( y \right) = \exp \left( - \left( 1 + \zeta y \right) ^ {- 1/\zeta} \right), \hspace{.2cm} \left( 1 + \zeta y \right) ^ {- 1/\zeta} > 0 
\end{equation}
for $y = ax + b$ with $a,\,b\in\mathbb{R}$ and $a>0$.
The Fisher-Tippett-Gnedenko theorem (Thm.~1.1.3 in~\onlinecite{Haan_EVT}) 
states that this single-parameter family, called generalized extreme value distribution (GEV), includes all
possible limiting distributions for the extreme values of a sample of i.i.d.~random variables.
The family contains three subclasses which exhibit quite different behavior, the Gumbel ($\zeta=0$), Fr\'echet ($\zeta>0$) and Weibull ($\zeta<0$) distributions. Qualitatively, the deciding criterion for a PDF to belong to the basin of attraction of one family or another (i.e. for the extrema
of the samples to be described by the corresponding family) is the form of its tails.\cite{Haan_EVT}

In our particular problem, as we try to find the minimum $\lambda_M$ for a certain configuration, we are effectively sampling from the left tail of $p(\Lambda_{M})$, which is the distribution of eigenvalues $\Lambda_{M}$ of $\mathcal{H}_{\mathrm{eff}}^{(M)}$. Typically the eigenvalues of a matrix are not uncorrelated, and thus GEV is \textit{a priori} not expected to describe the probability distribution of extreme eigenvalues\cite{Biroli_2007,astrauskas16}. Nevertheless, our results indicate that the distribution \eqref{eq:gev} provides indeed a very good description for our data, with the particular form depending only 
on the asymptotic behavior of $p(\Lambda_{M})$ for small $\Lambda_M$.

\section{ Structure of the slow operators} 
\label{sec:structure}

We will first study the structure of the slowest operators. In the strong disorder regime, we expect that the slow operators correspond to some LIOMs, or more precisely to a truncated version of them since our support size is fixed.
It is worth noticing here that although the slow operator method does not directly target the (truncated) LIOMs, as was done in Refs. \onlinecite{Kulshreshtha_2017, Goihl_2017}, in the localized regime, truncated LIOMs and their combinations are good candidates to attain (exponentially) small commutators.
\begin{figure} 
	
	\hspace{-.5cm}

		\subfloat{
		\begin{picture}(0,0)
		\put(60,93){\includegraphics[scale=.25]{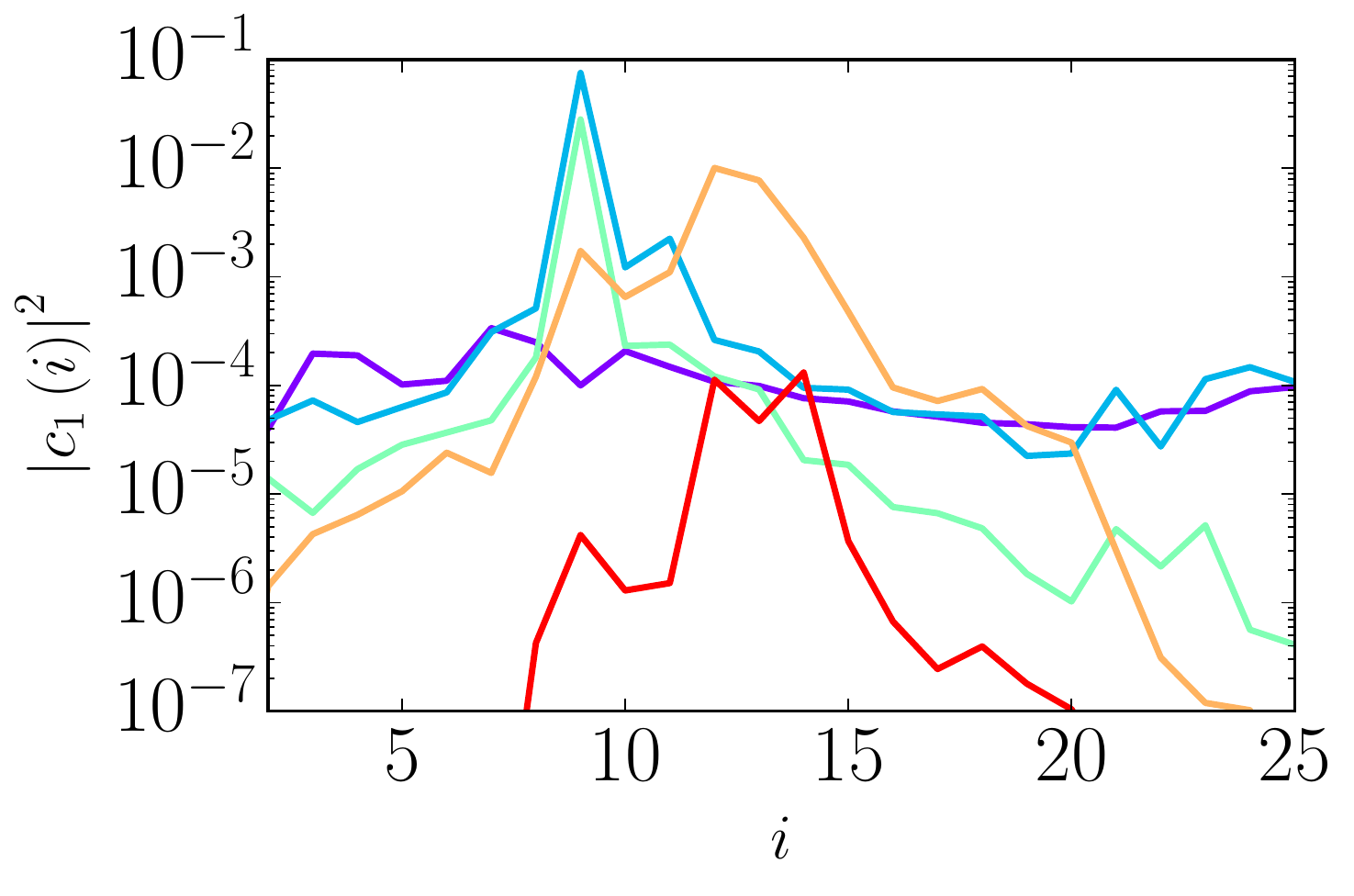}}
		\end{picture} 
		\includegraphics[scale=.60]{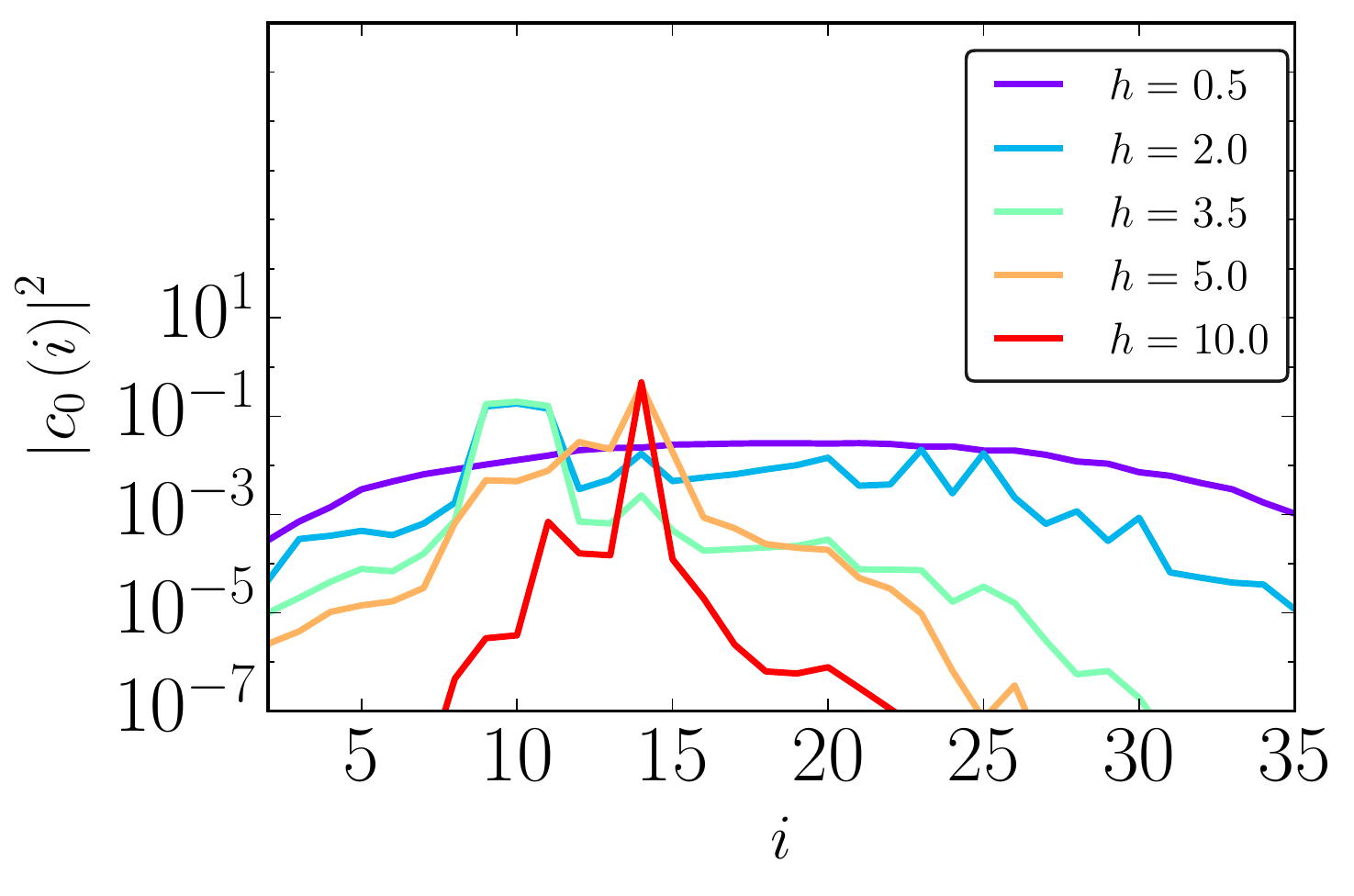}
	}

	\caption{\textbf{Local structure of the slowest operator with support $M=40$.}  The weight of single-site $\left| c_0 \left( i \right) \right|^2$ contribution is shown as a function of the position $i$ for various disorder strengths for a single disorder realization. At weak disorder, the local contributions are spread over the whole support and has the shape of a sine as expected from diffusive dynamics. For strong disorder, the operator becomes increasingly localized and exhibits exponential tails. The optimization has been performed for MPO bond dimension $D=100$. \textbf{Inset:} For larger support contributions such as $\left| c_1 \left( i \right) \right|^2$, we observe similar tails with considerably smaller values.}
	\label{fig:op_prof}
\end{figure}

\begin{figure*}
	\subfloat{
		\begin{picture}(0,0)
		\put(123,78){\includegraphics[scale=.29]{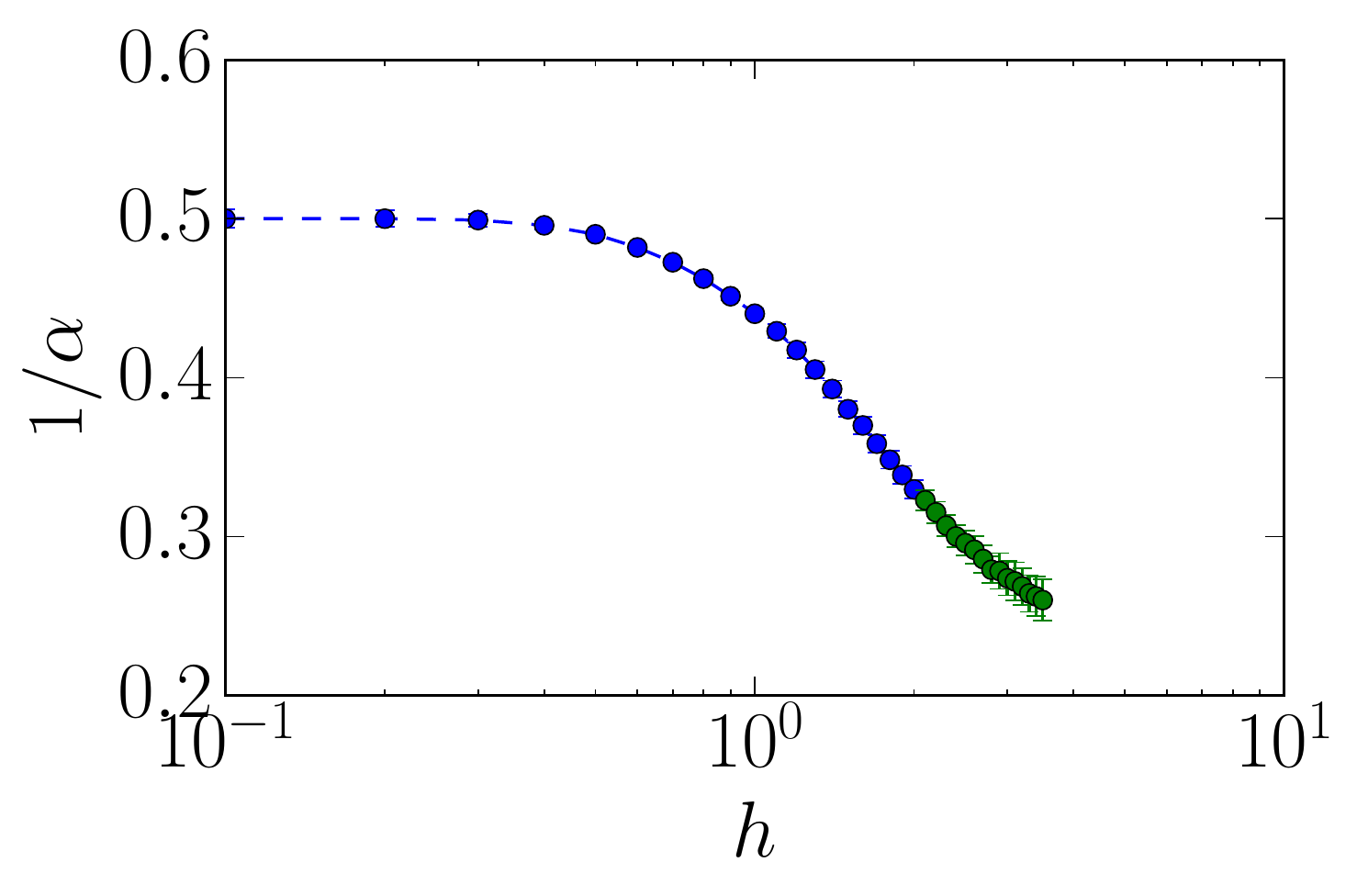}}
		\end{picture} 
		\includegraphics[scale=.58]{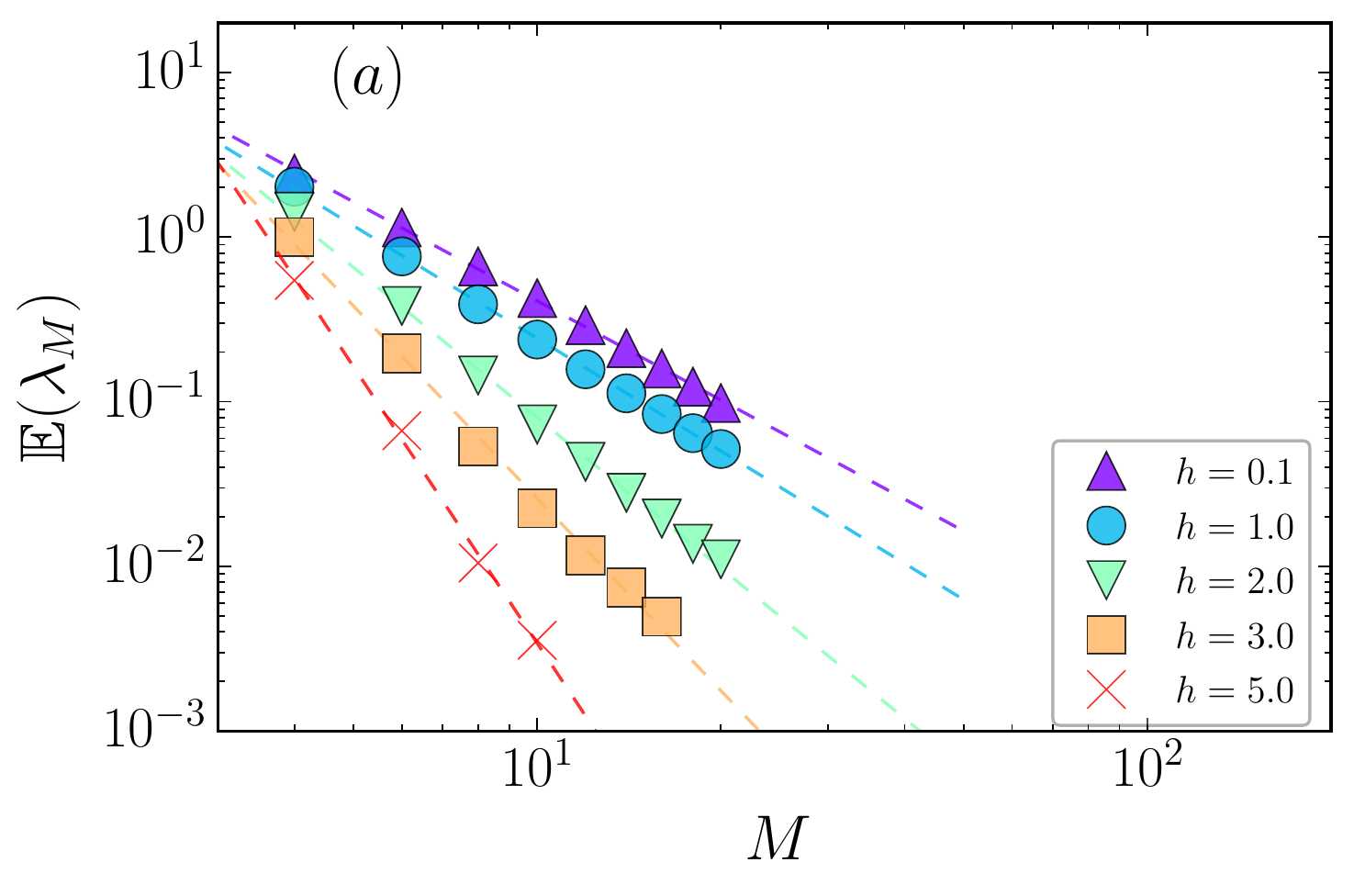}
	}
	\subfloat{  
		\begin{picture}(0,0)
		\put(117,80){\includegraphics[scale=.29]{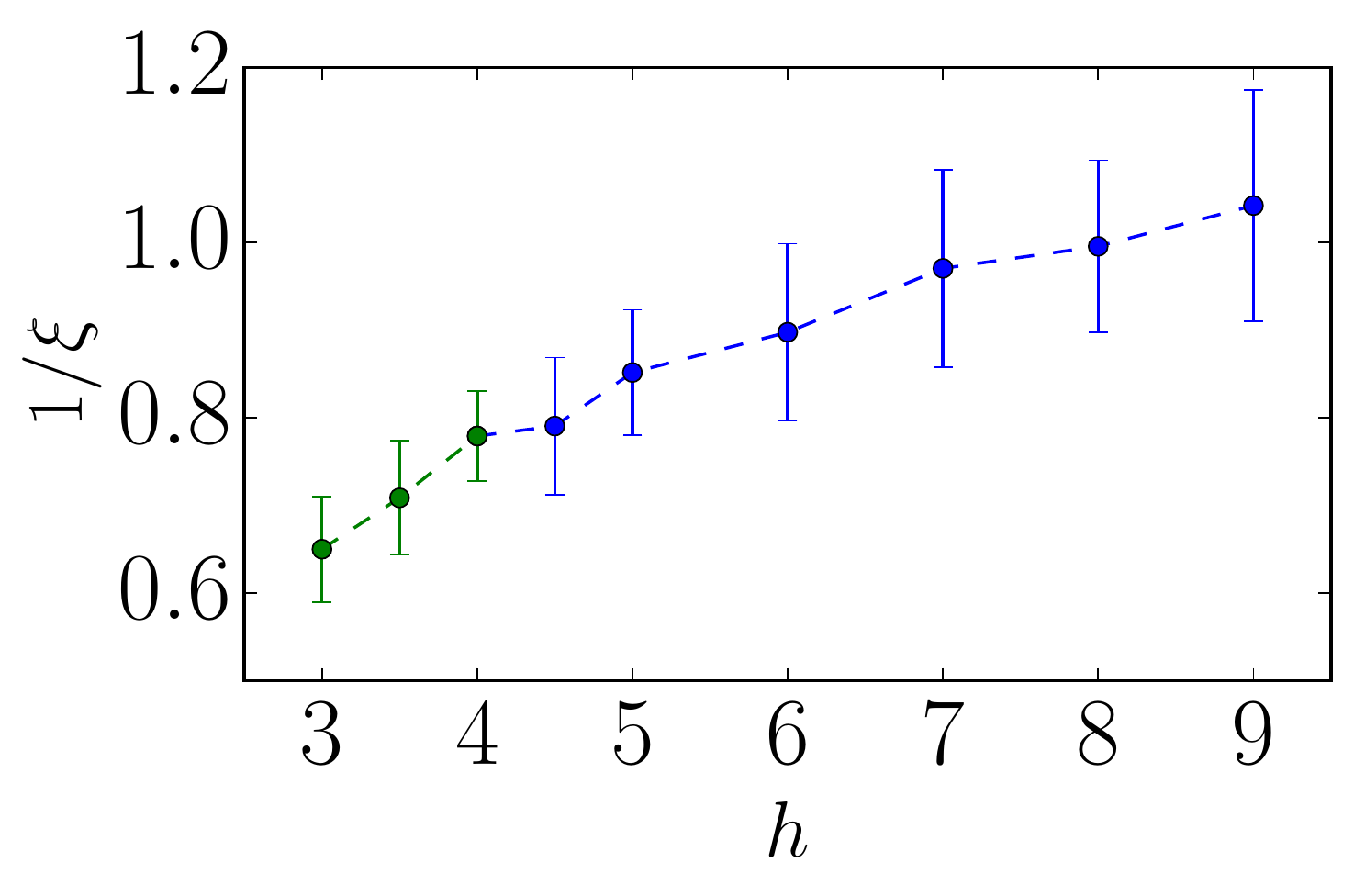}}
		\end{picture} 
		\includegraphics[scale=.58]{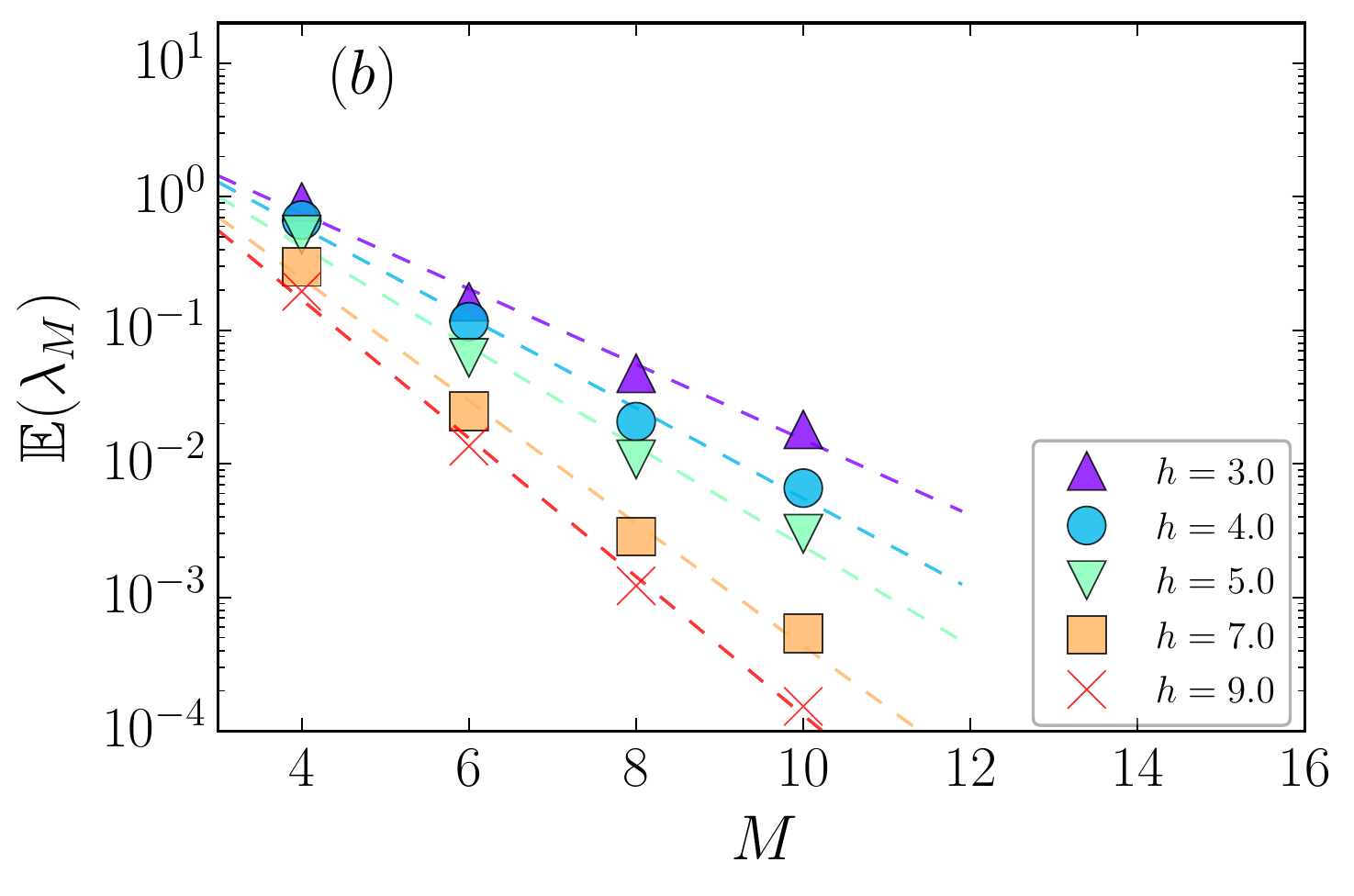}
	}
	\caption{\textbf{Disorder average of the smallest commutator $\mathbb{E}(\lambda_M)$.} \textbf{(a)} Decay of $\mathbb{E}(\lambda_M)$ at weak disorder. We observe that the average of $\lambda_M$ scales polynomially $M^{-\alpha}$ with an exponent that increases as the transition is approached. \textbf{Inset:} The inverse exponent $1/\alpha$ as a function of the disorder strength $h$ is shown. \textbf{(b)} Decay of $\mathbb{E}(\lambda_M)$ at strong disorder. For strong disorder, the decay of $\lambda_M$  is compatible with an exponential form, $e^{-M/\xi}$. \textbf{Inset:} 
	The inverse length scale $1/\xi$ decays as the transition point is approached. }
	\label{fig:lambda_thermal}
\end{figure*}

To address this question, we can analyze to which extent the slowest operators are local by examining their spatial structure in different regimes.
This can be done by studying their decomposition as a sum of tensor products of single-site Pauli matrices, $O_M= 2^{-M/2}  \sum_{\{\alpha_j\}=0}^3 C_{\alpha_1\ldots\alpha_M} \sigma_1^{\alpha_1}\ldots \sigma_M^{\alpha_M}$.
In principle, for an operator $O_M$ found by the minimization, either exactly or as a MPO approximation, we can efficiently evaluate any single coefficient  $C_{\alpha_1\ldots\alpha_M} =2^{-M/2} \mathrm{tr}(\sigma_1^{\alpha_1}\ldots \sigma_M^{\alpha_M} O_M)$. Nevertheless, due to the exponential growth of the basis dimension with the support, already for moderate values of $M$ it is unfeasible to inspect all the individual coefficients. Instead, a more physical quantity for exploring the localized nature of the operators is the combined weight of all terms with a fixed range which are supported on a certain subregion. 
We define a range-$k$ operator as a product of Pauli matrices that act non trivially on sites $i$ to $i+k$. Formally, this corresponds to the operator $ \Theta _{i, k}^{ a,b, \{ \alpha_m\}}  = \sigma^0_1 \ldots  \sigma^0_{i-1} \sigma_i^{a} \sigma_{i+1}^{\alpha_1} \ldots \sigma_{i+k-1}^{\alpha_{k-1}}\sigma_{i+k}^{b} \sigma^0_{i+k+1} \ldots  \sigma^0_{M}$, where $a, b \in 1,2,3 $ to ensure Pauli matrices at the edges (thus imposing that the support is strictly $i \rightarrow i +k$) and $ \{ \alpha_m\} \in 0,1,2,3$.

The weight of these range-$k$ terms can be written as 
\begin{align} \label{eq:k-range_operator}
|c_k(i)|^2 := \frac{1}{2^{M}} \sum_{a,b=1}^3 \sum_{\{\alpha_m\}=0}^3 \left| \text{Tr} \left( \Theta _{i, k}^{ a,b, \{ \alpha_m\}} O_M \right) \right|^2.
\end{align}

We thus solve the optimization \eqref{lambda} and compute the weights, $|c_k(i)|^2$, along the chain of terms with fixed range $k$. Even if the support is large, as $O_M$ is written as an MPO, this quantity can be computed  efficiently, i.e., with a cost that only scales polynomially in the range $k$ and the support $M$.\footnote{The reason for this efficient computation is that the weight $|c_i(k)|^2$ can be expressed in terms of the vectorized operator $O_M$ as the expectation value of a superoperator with tensor product structure and can then be evaluated with MPS primitives.}

We find clear differences in the structure of the operator depending on the disorder strength. This is illustrated in Fig. \ref{fig:op_prof} for the MPO, that minimizes Eq. \eqref{lambda} for a particular disorder realization $h_i = h \cdot r_i$, where $r_i$ are fixed random numbers for each value of $h$. We have chosen a support of $M=40$ sites and a relatively large bond dimension of $D=100$, in order to ensure that truncation errors are negligible compared to the effects we observe. The figure shows the different spatial profile of the single-site contributions $\left| c_0 \left( i \right) \right|^2 = 2^{-M} \sum_{a=1}^3  |\text{Tr} \left( \sigma_i^{a} O_M \right)| ^2$ and two-site contributions $\left| c_1 \left( i \right) \right|^2$ (Eq.\eqref{eq:k-range_operator} with $k=1$)  for different values of the disorder strength.  For strong disorder, where the LIOMs are exponentially localized, at least one of the $\left| c_0 \left( i \right) \right|^2$ is expected to be dominant. Indeed, we observe that for large disorder the operator is well localized around a single site, 
with weights that decay exponentially around this point.
As the disorder strength decreases, the weights decay slower with distance, and the tails are no longer exponential. Finally for very small $h$ the profile becomes flat and has the shape of a sine as expected in the diffusive regime. The single-site terms shown in the figure always accumulate most of the weight.
We found that higher order terms Fig. \ref{fig:op_prof} (inset), which have smaller contributions, exhibit a similar spatial decay. 
The observed profile at strong disorder  is in good agreement with what one could expect from the decomposition 
of the LIOMs in the canonical operator basis, what would in principle allow the extraction of a localization length scale of the LIOM.

\section{Average commutator}\label{sec:avg}

The precise value of the minimum $\lambda_M$ for fixed support $M$ and disorder strength $h$ will depend on the disorder realization. Hence, the average over realizations $\mathbb{E}(\lambda_M)$ will provide information about the underlying phase and also indirectly about transport. 

In the following, we will optimize Eq. \eqref{lambda} over the family of MPOs with finite bond dimension $D=10$. This allows us to reach large support sizes $M$ and collect a significant amount of statistics. The average of the smallest commutator $\mathbb{E}(\lambda_M)$ is shown as a function of the support size $M$ for different disorder strength $h$ in Fig. \ref{fig:lambda_thermal}. At very weak disorder $h \lesssim 0.4$ our data recovers the diffusive scaling $\mathbb{E}(\lambda_M) \sim M^{-2}$ of conserved quantities. 

When increasing the disorder strength, the decay remains a powerlaw $\mathbb{E}(\lambda_M) \sim M^{-\alpha}$, however, with an increasing exponent $\alpha$ indicating that the transport becomes slower due to the presence of disorder, see inset of Fig. \ref{fig:lambda_thermal} (a) which shows the inverse exponent $1/\alpha$. This is consistent with the observation of sub-diffusive transport on the thermal side of the MBL transition.\cite{Agarwal, BarLev_Absence_2015, znidaric_diffusive_2016-1} The longer relaxation times in this regime can be interpreted as follows: rare insulating regions with larger than typical disorder
act as bottlenecks for the conventional diffusive transport.\cite{Agarwal, VHA, PVPtransition} Upon approaching the MBL transition at $h \sim 4$,~\cite{Luitz_2015} we expect the exponent to diverge. Our data indeed shows that the power law decay becomes very steep. Due to the finite support sizes $M$ and due to the slow convergence of the algorithm for large disorder, we, however, find that the exponent levels off at a finite value. For practical purposes, we adopt the strategy of limiting the maximum computational effort for each search. Therefore, the result of the algorithm provides an upper bound for the true minimum $\lambda_M(h)$, and the corresponding $\alpha$ is biased towards smaller values.

When crossing the MBL transition, an extensive number of exponentially localized conserved quantities, corresponding to the LIOMs $\tau_i^z$ in Eq. \eqref{eq:lbits}, emerges.\footnote{There will actually be exponentially many conserved quantities, which can be constructed by any product or combination of LIOMs $\tau_i^{z}$.} Therefore, provided the window $M$ is sufficiently large to capture most of the support of any LIOM, the smallest commutator should decay exponentially as $\lambda_M \sim e^{-M/\xi}$. Such an exponential decay is supported by our numerical simulations for strong disorder, see Fig. \ref{fig:lambda_thermal} (b). The length scale $\xi$ is related (in a possibly non-trivial way) to the localization length of the LIOMs. In the inset, we show that with increasing disorder $h$, the inverse length scale $1/\xi$ increases monotonically. The decays of $\alpha^{-1}$ and $\xi^{-1}$ as the transition is approached are both compatible with previous estimates of the critical region, see e.g. Refs. \onlinecite{Luitz_2015,Kulshreshtha_2017}.

\section{Extreme value theory} 
\label{sec:evt}

\subsection{Numerical results}

\begin{figure}
\includegraphics[scale=.5]{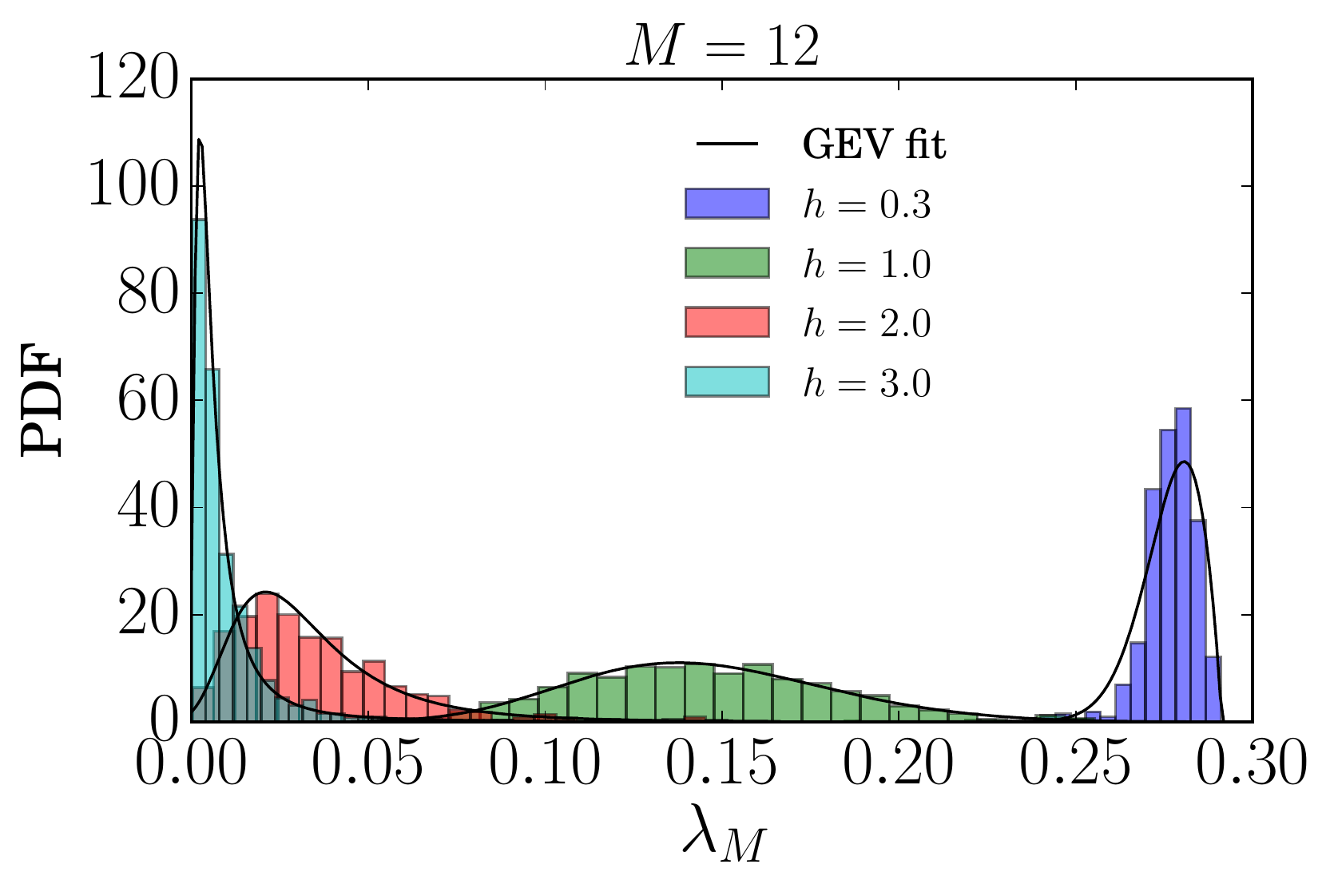}
\caption{\textbf{Probability density function of $\lambda_M$.} Using the maximum likelihood method we fit the numerical data for $\lambda_M$ with the GEV distribution \eqref{eq:gev}. The shape of the distribution function is strongly influenced by the presence of disorder.
}
\label{fig:fit_hist}
\end{figure}

\begin{figure*}
\centering
\subfloat{\includegraphics[scale=.57]{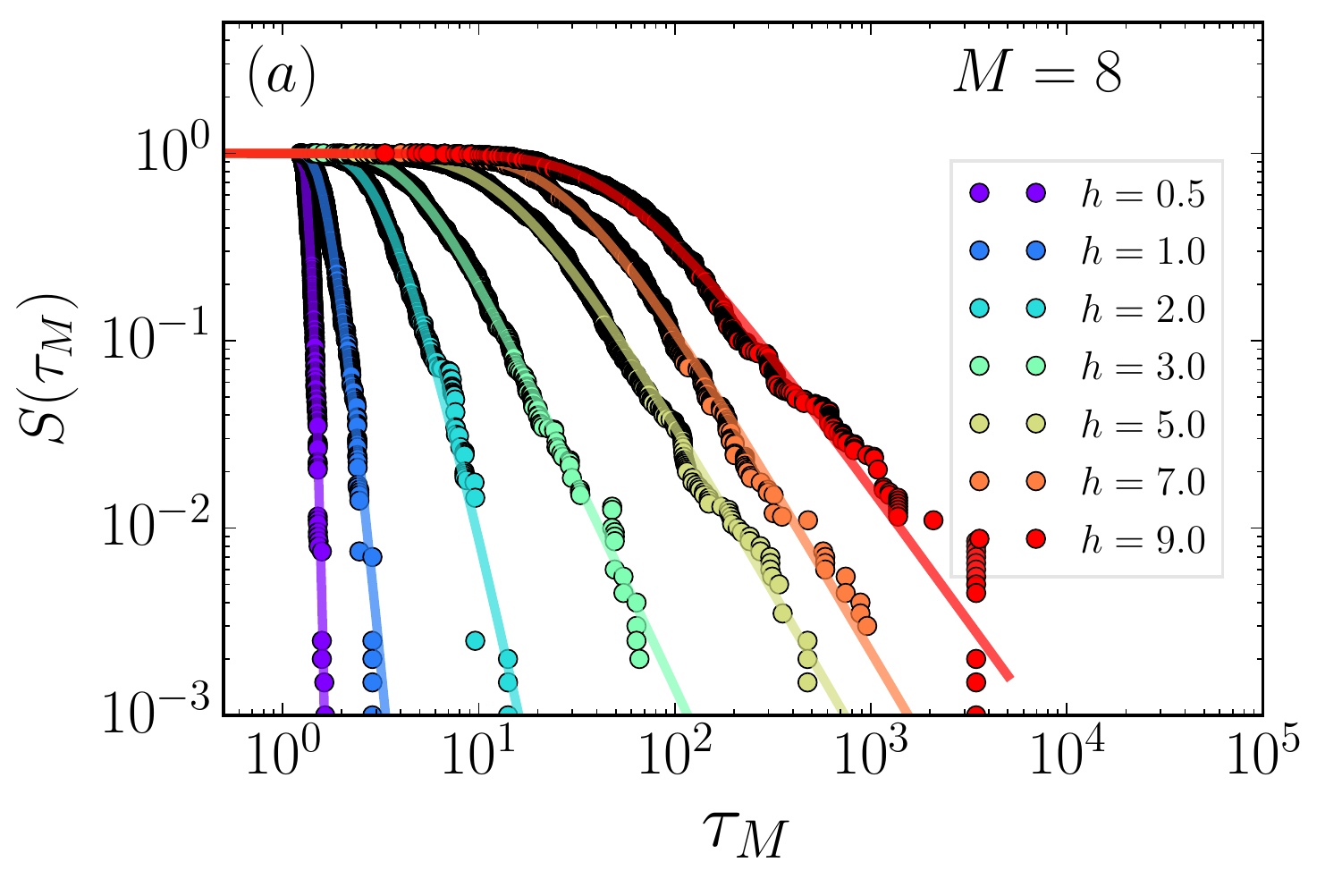}}
\subfloat{\includegraphics[scale=.57]{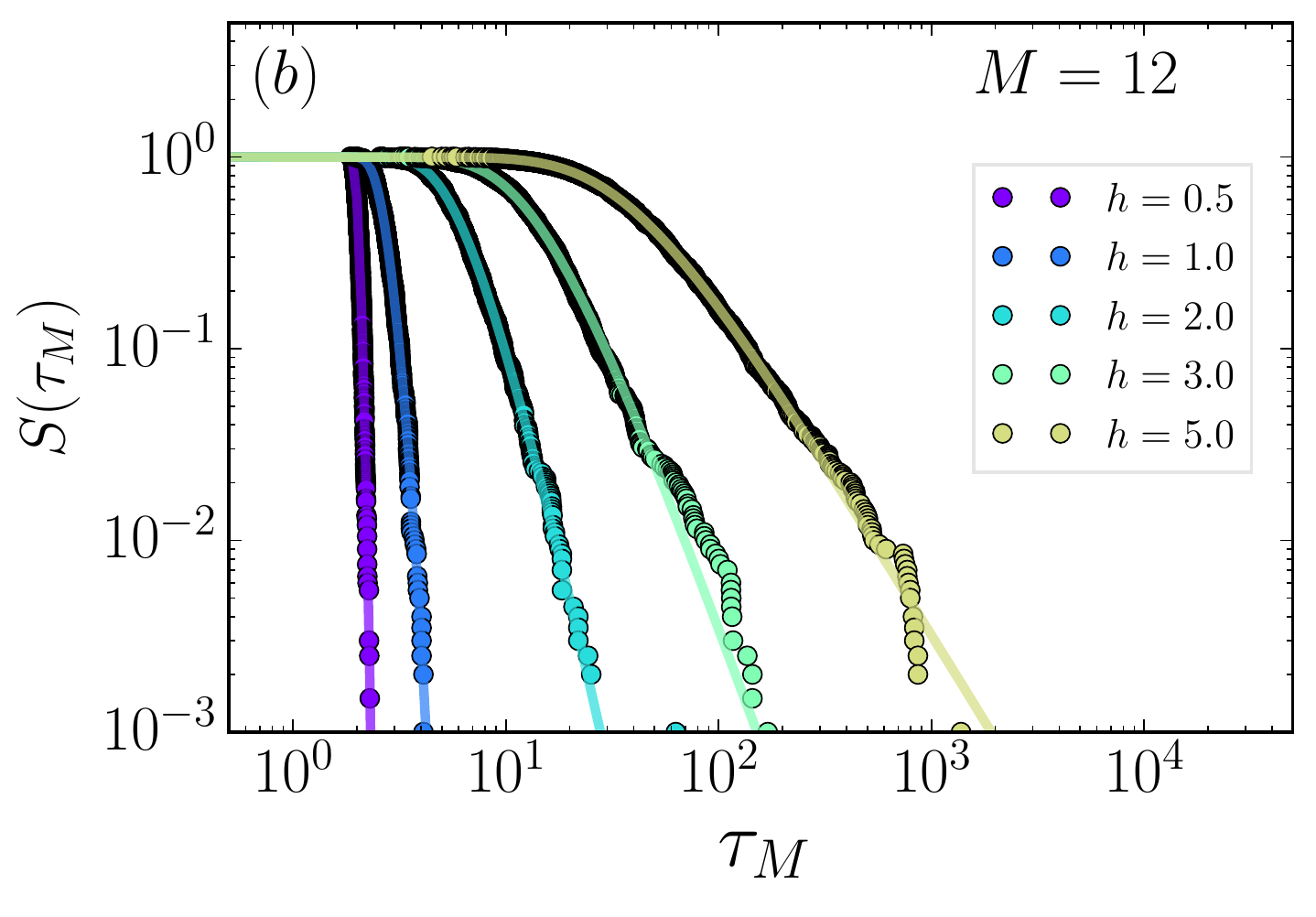}}
  \caption{\textbf{Survival function of $\tau_M = \sqrt{1/\lambda_M}$.} We fit the survival function of $\tau_M$ to the GEV, Eq. \eqref{eq:gev}, which decays super-exponentially for weak disorder and polynomially for large disorder. The form of these tails is sensitive to rare Griffiths effects. }  
\label{fig:fit_tails}
\end{figure*}

We now study the probability distributions of the slow commutators $\lambda_M$, which can be sensitive to rare events. In particular, we show that EVT provides useful mathematical tools to characterize the tails of the probability density and cumulative distribution functions.
We first illustrate in Fig. \ref{fig:fit_hist}, that the GEV distribution, obtained from differentiating Eq. \eqref{eq:gev}, provides a very good fit to the collected data over a range of disorder strengths. Moreover, we find that the shape of the distribution strongly depends on the disorder value and that the peak of the distribution shifts from a finite, relatively large value for weak disorder, toward zero as $h$ increases. In between those limits, in a regime where one expects sub-diffusive dynamics, the distribution broadens considerably. 

In the following we will analyze the tails of this distribution quantitatively. 
To this end, we introduce a new variable, with dimensions of time, as $T_M=(\Lambda_M)^{-1/2}$. The minimal eigenvalue $\lambda_M$ thus corresponds to the maximum of $T_M$, that we call $\tau_M$, and has the physical interpretation of bounding the thermalization time of the operator $\rho(O_M)$ from below. This transformation of variables moves the left tails of $p(\lambda_M)$ that arise due to rare events, to right tails of $p(\tau_M)$. The corresponding survival functions $S(\tau_M) = 1-F(\tau_M)$ are shown in Fig. \ref{fig:fit_tails} on logarithmic scale. We observe that for small disorder, the survival function approaches zero super-exponentially fast describing weak tails, due to the small probability of large $\tau_M$. As the disorder increases, the survival function decays with a power-law tail. The value of the disorder at which the tails change is in agreement with the crossover from diffusive to sub-diffusive dynamics. 

To make this observation more quantitative, we study the variation of the shape parameter $\zeta$ from the fitted GEV, defined in Eq. \eqref{eq:gev}, with the disorder strength. Our data shows that $\zeta$ shifts from values close to zero for weak disorder, to clearly positive values as the MBL phase is approached. The $\zeta$ parameter determines the type of distribution. In particular, the deviation of $\zeta$ from zero describes the crossover from a peaked distribution to one with polynomial tails. In terms of the GEV families, this corresponds to a change from a Gumbel ($\zeta = 0$) to a Fr\'echet ($\zeta > 0$) distribution. 
The observed qualitative change can be explained by an intermediate regime in which atypically slow operators appear for any given support $M$, leading to strong tails in the probability density function of $\tau_M$. The value of the disorder strength at which the shape parameter $\zeta$ starts being clearly positive is again consistent with the expected sub-diffusive regime of the thermal phase.\cite{znidaric_diffusive_2016-1}

\subsection{Interpretation in terms of Griffiths effects}
\label{sec:thermalEVT}

Although the eigenvalues of the effective operators in our study are not expected to be uncorrelated, the results in the previous section show that GEV does indeed describe our findings accurately. Hence, in this section we use EVT arguments to show how
the observed distributions can be qualitatively explained in terms of the existence of rare Griffiths regions.

We first consider the effect of rare localized regions in the thermal phase but close enough to the transition. In this situation, given a support size $M$, the \emph{typical} values of $\lambda_M$ will be polynomial in $M$. Nevertheless, if rare regions are present that (partly) support an exponentially localized operator, they can give rise to exponentially small values of $\lambda_M$. More concretely, within a fixed window $M$, 
such a rare configuration of size $\ell < M$ will occur with an exponentially small probability $p(\ell)\sim c^{\ell}$, for some $c<1$. This patch can support a localized operator, with localization length $\xi$, with a correspondingly small commutator $\Lambda \sim e^{-\ell/\xi}$ (or equivalently\citep{Juhasz_2006} $T\sim e^{\ell / 2 \xi}$).
Strictly speaking, this will be detected as the slowest operator only if this commutator is smaller than the (polynomial) typical commutator for the complementary region of size $M-\ell$. Since we are interested in the probability for very small commutators,
and there is an exponential separation between the scaling of both terms,  
we may assume that small enough commutators will always come from such rare patches. 
Evidently, this can only be true for commutators below a certain ($M$ and $h$ dependent) threshold. With this assumption, the probability of some very small value of $\Lambda$ is determined by the probability of finding a rare patch such that  $\Lambda \sim e^{-\ell/\xi}$, and thus we can identify $p( \ell ) |d \ell| = p\left( T \right) |dT|$. From that we find that, for the range of $T$ that correspond to rare regions of length $\ell < M$, the probability for the largest $T$ values is polynomial,
\begin{equation}\label{poly_tails}
p\left( T \right) \sim 2 \xi T^{- 2 \xi \left| \ln\left(c\right) \right| -1 }.
\end{equation} 
For PDFs with polynomial tails, EVT predicts that the extreme values will be governed by a Fr\'echet distribution with $\zeta>0$ (see appendix \ref{app:EVT}).

\begin{figure}
	\includegraphics[scale=0.55]{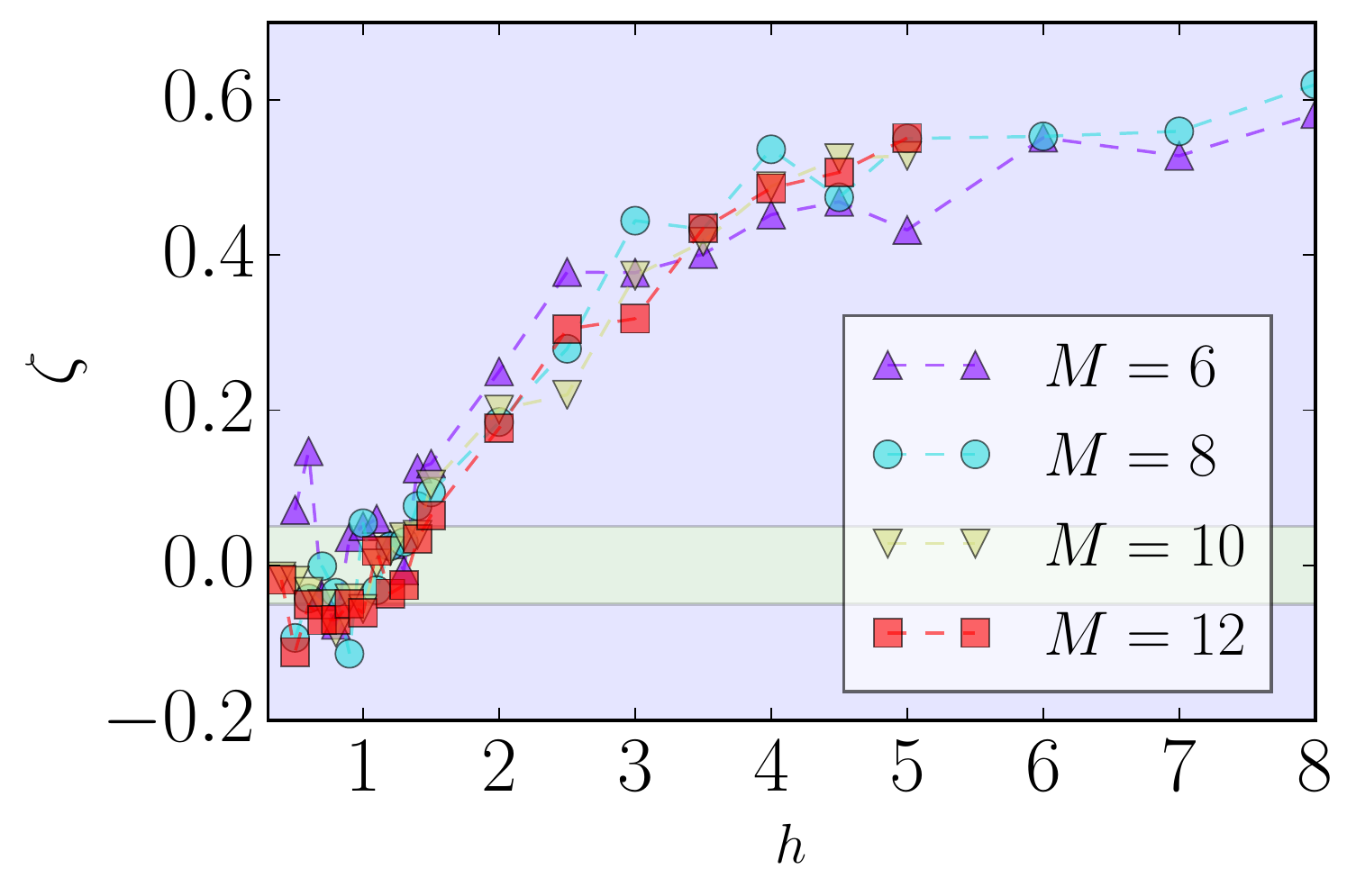}
	
	\caption{\textbf{Shape parameter $\zeta$ of the generalized extreme value distribution as a function of the disorder strength.} The values of the shape parameter $\zeta$ range from $\sim 0$ (corresponding to a Gumbel distribution) to $\zeta \sim  1/2$ (corresponding to Fr\'echet distribution) as the disorder $h$ is increased.}
	\label{fig:shape_param}
\end{figure}

For very weak disorder, instead, the same rare configurations of the field will not support an exponentially localized operator, so that, even if a large effective disorder on some region may give rise to commutators that are below the typical one, 
they still decay at most polynomially. Consequently, the argument above does not apply and 
we expect the probability of the smallest values to decay faster (even exponentially).
For a PDF with such properties, EVT predicts a Gumbel distribution with $\zeta = 0$ (see appendix \ref{app:EVT}). This is in agreement with the shape parameter $\zeta$ shown in Fig. \ref{fig:shape_param} where, for weak disorder, we obtain values of $\zeta$ very close to zero which correspond to a Gumbel distribution.

Beyond the localization transition we expect the typical regions to (partly) support localized operators, and thus give rise to exponentially small values of $\lambda_M$ (correspondingly exponentially large values of $\tau_M$), as we explicitly observed in the average values shown in Section~\ref{sec:avg}. 
In our data for strong disorder, we have also found broad tails of the PDF $p(T)$ which are consistent with a power law. Thus, the function describing our data resembles the Fr\'echet distribution. This might be explained by the fact that matrix elements in the MBL phase have been found to exhibit broad tails.~\cite{Gopalakrishnan15,pekker_fixed_2017}

As we approach the transition from the MBL side by lowering the disorder, rare thermal inclusions can appear that potentially correspond to larger than typical commutators. Yet, our method only looks for the smallest commutator in each given window. Thus,
if a support $M$ encloses one such thermal subregion, a competition arises between the values of the commutator for the inclusion and that of the (typically) exponentially localized complement. Because we expect that a thermal inclusion gives rise to only polynomially decaying commutators, their value can only be the smallest one when the inclusion is sufficiently %dominating the support of size $M$. 
large in relation to the size M of the support.
This causes our method to be less sensitive to rare thermal regions in the localized side of the transition. 

In contrast to other numerical studies, where the presence of Griffiths effects was inferred from averaged observables,\cite{Agarwal, BarLev_Absence_2015, znidaric_diffusive_2016-1} with our method we may directly locate rare regions. In particular, we can obtain the disorder potential for the eigenvalues $\lambda_M$ that contribute to the tail of the distribution and analyze the microscopic configuration of the random field  $\{h_i\}$ in real space. Following this procedure, we could, however, not unambiguously determine an obvious correlation between strong fluctuations of the field in real space and small commutators. It remains an open question whether it is possible to predict the location of the rare regions from the disorder landscape using a more direct method than the optimization.

\section{Discussion}\label{sec:discussion}
We have constructed slow operators with finite support by minimizing their commutator with the Hamiltonian of the system using both exact diagonalization and tensor network techniques. In particular, we have considered the Heisenberg spin chain with random magnetic field, which displays a dynamical transition from the thermal to the many-body localized phase. The scaling of the minimal commutator with support size provides information on the localization transition as well as on transport in the system without resorting to a specific initial state. 

Furthermore, we have demonstrated that the tails of the probability distributions are sensitive to rare insulating regions in the thermal phase near the many-body localization transition.
We have shown that the statistics of the smallest commutators can be analyzed
within the mathematical framework of extreme value theory.~\cite{Haan_EVT} In particular, we have found that the distributions are well described by generalized extreme value functions whose shape depends on the disorder strength. By extreme value theory arguments, the observed behavior in the tails can be connected to the appearance of rare, strongly disordered regions, that give rise to atypically small minimal commutators. In particular, the disorder strength at which the distribution functions obtain power-law tails is consistent with the appearance of sub-diffusive transport.~\cite{Agarwal,BarLev_Absence_2015, znidaric_diffusive_2016-1}

We conclude that the slow operator technique combined with extreme value theory, constitutes a valuable tool for exploring microscopic mechanisms of the MBL transition. Further developments may include tailoring the optimization technique to target explicitly the rare thermal regions on the localized side, which could provide new insights into this less explored aspect of MBL physics.  Another intriguing question would be how coupling an MBL system to an external bath~\cite{Nandkishore14,fischer2016dynamics,levi2016robustness,medvedyeva2016influence,gopalakrishnan_noise-induced_2017} would change the structure of slow operators.

\acknowledgements
We acknowledge financial support from ExQM and IMPRS (N.P.) and from the Technical University of Munich - Institute for Advanced Study, funded by the German Excellence Initiative and the European Union FP7 under grant agreement 291763 (M.K.). This work was also partially funded by the European Union through the ERC grant QUENOCOBA, ERC-2016-ADG (Grant no. 742102).

\begin{appendix}
\section{Extreme value theory}
 \label{app:EVT}

A given PDF is said to belong to the basin of attraction of one of the extreme values distributions,
namely Gumbel, Fr\'echet or Weibull, when the extrema are distributed according to the corresponding function.
The von Mises conditions\cite{Haan_EVT} establish simple criteria to determine 
whether $p(x)$ belongs to one of them.
In this appendix we show how the conditions apply to the particular cases of the distributions discussed in 
section \ref{sec:thermalEVT}.

\paragraph{Strong disorder implies a Fr\'echet distribution.}

Rare regions in the thermal phase near the MBL transition may support localized operators. As shown in Eq. \eqref{poly_tails}, the corresponding probability distribution
is expected to decay as $p(T)\sim 2 \xi T^{-2 \xi |\ln c\,| - 1}$. A sufficient condition\cite{Haan_EVT} for a PDF $p(T)$ to belong to the basin of attraction 
of the Fr\'echet distribution is that the corresponding CDF $F (T) = \int_{- \infty} ^T p(T') dT'$ satisfies the condition
\begin{equation}
\lim_{T \to \infty}  \frac{ T   F'(T)}{1 - F(T) }  =  \frac{1}{\zeta},
\label{eq:Weibullcond}
\end{equation}
with $\zeta > 0$.

From Eq. \eqref{poly_tails} we obtain
\begin{equation}
\frac{ p(T)}{1-F(T)}=\frac{p(T)}{\int_T^{\infty}p(t)dt}=\frac{2\xi |\ln c\, |}{T},
\end{equation}
so that
\begin{equation}
\frac{ p(T)}{1-F(T)}=\frac{2\xi |\ln c\, |}{T},
\end{equation}
and, asymptotically,
\begin{equation}
\lim_{T\to\infty}\frac{ T p(T)}{1-F(T)}= 2\xi |\ln c\, |>0,
\end{equation}
which ensures the condition above with $\zeta^{-1}=2\xi |\ln c\, |$,
and thus implies a limiting Fr\'echet distribution.

\paragraph{Exponentially decaying tails imply a Gumbel distribution.}

In order to prove that a PDF belongs to the basin of attraction of the Gumbel distribution it is sufficient to check the following condition \cite{Haan_EVT}
\begin{equation}
\lim_{T \uparrow T_{max}} \frac{d}{dT} \left( \frac{1 - F (T) }{F'(T)} \right) =0,
\label{eq:GEVcond}
\end{equation}

We assume the simplest exponential decay for the right tail of the distribution $p(T) \sim e^{-kT}$.
From the corresponding cumulative function we get the survival probability,
\begin{equation}
1-F(T)=\int_T^{\infty}p(t)dt \propto \frac{1}{k}e^{- k T},
\end{equation}
so that
\begin{equation}
\frac{1-F(T)}{p(T)}\sim\frac{1}{k}.
\label{eq:ratio}
\end{equation}
Thus, the derivative vanishes, which ensures Eq.\eqref{eq:GEVcond}. 

\end{appendix}

\bibliography{library}

\end{document}